\pdfoutput=1
\documentclass[12pt,letterpaper,titlepage,reqno,oneside]{article}
\usepackage[utf8]{inputenc}
\usepackage{doi}
\usepackage{graphicx}
\usepackage[super,comma]{natbib}
\usepackage[english]{babel}
\usepackage[figurename=FIG.,tablename=TABLE]{caption}
\usepackage{enumerate}
\usepackage{amsmath,amsfonts,amssymb,wasysym,latexsym,amsthm,bm}
\usepackage[mathscr]{eucal}
\usepackage{color}
\usepackage{subfig}
\usepackage{siunitx}
\usepackage{textcomp}
\usepackage{titlesec}
\usepackage{multirow}
\usepackage[subfigure,titles]{tocloft}
\usepackage{setspace}

\DeclareCaptionFormat{upper}{#1#2\uppercase{#3}\par}

\titleformat{\section}{\normalsize\bfseries}{\Roman{section}.}{1em}{\normalsize\uppercase}
\titleformat{\subsection}{\normalsize\bfseries}{\Alph{subsection}.}{1em}{}
\titleformat{\subsubsection}{\normalsize\itshape}{\arabic{subsubsection}.}{1em}{}

\renewcommand{\thesection}{\Roman{section}}

\cftpagenumbersoff{figure}
\cftpagenumbersoff{subfigure}

\makeatletter
\renewcommand\listoffigures{%
    \medskip\raggedright \textbf{LIST OF FIGURES}%
    \@mkboth{\MakeUppercase\listfigurename}%
        {\MakeUppercase\listfigurename}%
    \@starttoc{lof}%
}
\makeatother

\makeatletter
\def\p@subsection{\thesection.\,}
\makeatother

\begin{document}
\sisetup{detect-family,detect-display-math=true,exponent-product=\cdot,output-complex-root=\text{\ensuremath{i}},per-mode=symbol}

\author{%
Michael A.G. Timmer\footnote{Author to whom correspondence should be addressed. E-mail: timmer.mag@gmail.com} and Theo H. van der Meer \\
\textit{Department of Thermal Engineering, University of Twente,}\\
\textit{Enschede, The Netherlands}\\ \\
Kees de Blok\\
\textit{Aster Thermoacoustics, Veessen, The Netherlands}%
}
\title{\LARGE{Review on the conversion of thermoacoustic power into electricity} \\
\textit{\large{Review thermoacoustic to electric conversion}}
}
\date{\today}

\maketitle

\begin{abstract}
Thermoacoustic engines convert heat energy into high amplitude acoustic waves and subsequently into electric power. This article provides a review of the four main methods to convert the (thermo)acoustic power into electricity. First, loudspeakers and linear alternators are discussed in a section on electromagnetic devices. This is followed by sections on piezoelectric transducers, magnetohydrodynamic generators, and bidirectional turbines. Each segment provides a literature review of the given technology for the field of thermoacoustics, focusing on possible configurations, operating characteristics, output performance, and analytical and numerical methods to study the devices. This information is used as an input to discuss the performance and feasibility of each method, and to identify challenges that should be overcome for a more successful implementation in thermoacoustic engines. The work is concluded by a comparison of the four technologies, concentrating on the possible areas of application, the conversion efficiency, maximum electrical power output and more generally the suggested focus for future work in the field. \\

\noindent \textit{Copyright 2018 Acoustical Society of America. This article may be downloaded for personal use only. Any other use requires prior permission of the author and the Acoustical Society of America.} \\

\noindent \textit{The following article appeared in J. Acoust. Soc. Am 143(2) and the final version in a proper two-column format may be found at: \\ \url{http://scitation.aip.org/content/asa/journal/jasa/143/2/10.1121/1.5023395}} \\

\end{abstract}

\addtocounter{page}{2}

\section{Introduction\label{sec:intro}}
\noindent This review paper provides an overview of the current technologies for converting acoustic power into electricity. The focus is on the field of thermoacoustics, in which available heat is converted into high-energy acoustic waves that can subsequently be utilized. A short description of thermoacoustics is provided in Sec.~\ref{sec:intro:genthermoacoustics}. After this the focus of this review paper, converting (thermo)acoustic power into electricity, is introduced in Sec.~\ref{sec:intro:thermoacousticselectricity}. 

\subsection{General thermoacoustics \label{sec:intro:genthermoacoustics}}
\noindent Thermoacoustics encompasses the fields of thermodynamics and acoustics. The basis lies in the thermoacoustic effect, first described by Lord Rayleigh as follows: `If heat be given to the air at the moment of greatest condensation, or taken from it at the moment of greatest rarefaction, the vibration is encouraged'.\cite{Rayleigh1945} A thermoacoustic device utilizing this conversion of heat into acoustic work is often referred to as a prime mover. The reverse direction of converting acoustic work into heat is also possible and such devices are referred to as heat pumps (or refrigerators). 

Utilizing the thermoacoustic effect has resulted in the first thermoacoustic engine designs around the 1950's from Bell Telephone Laboratories.\cite{hartley1951electric,marrison1958heat} These engines converted a temperature gradient into standing waves using `singing pipes',\cite{Ceperly1979} and subsequently produced electricity from the acoustic power using an electromagnetic converter. Although promising due to the simple and reliable concepts, the overall efficiency was still unsatisfactory. Ceperley stated that the latter was mainly due to the fact that their engines were based on standing wave phasing, where an imperfect heat transfer has to be present to facilitate the necessary phasing between the pressure and particle velocity. \cite{Ceperly1979} As an alternative, he proposed to use devices working on traveling waves,\cite{ceperley1982resonant} in which the gas undergoes a cycle similar to the inherently efficient Stirling cycle.\cite{Walker} In the past few decades after the work of Ceperley, both the standing wave and traveling wave thermoacoustic devices have been further developed. A unifying perspective of both thermoacoustic branches with their underlying mathematics and working principles is given in the book of Swift. \cite{swift2017thermoacoustics}  

Thermoacoustic engines can be used wherever there is a sufficient heat source available. The onset temperature difference for the device to start producing power depends on the design and operating conditions (e.g. mean pressure and working gas), but it can be relatively low compared to other technologies. For example, a four-stage traveling wave engine has been shown to start producing acoustic power for a temperature difference as low as $\sim$$\SI{30}{K}$.\cite{DeBlok2010} This opens the market for thermoacoustic devices wherever there is such a relatively small temperature difference available. As long as done cost-effectively, this can either boost the efficiency of current systems having a stream of unused waste heat or utilize a (possibly sustainable) heat source for stand-alone thermoacoustic devices. Possible application areas for thermoacoustics include waste-heat recovery,\cite{Gardner2009,Yu2012b,Tijani2013,DeBlok2008} solar powered devices,\cite{DeBlok2008,Chen2000,Wu2012} and small low-cost applications for e.g. rural areas.\cite{Chen2013,Yu2010,Yu2011a,Garrett2013,Jaworski2013a,Abdoulla-Latiwish2017a} An overview of more applications and thermoacoustic devices can be found in the works of Garrett\cite{Garrett2004} and Jin.\cite{Jin2015}

Besides the low temperature difference required to operate, thermoacoustics has gotten increasing attention due to several other advantages it has over competing technologies. One of the most important characteristics is the need for no or, in case of producing electricity, few mechanically moving parts. Furthermore, any moving part is not situated near the high temperature region of the thermoacoustic engine, therewith reducing the material requirements when compared to conventional technologies such as automotive engines or Rankine cycle based power plants. The lack of moving parts and low material requirements makes thermoacoustic devices inherently simple, robust, and economic to produce and results in reliable devices that require little to no maintenance. As Ceperley stated, this lack of moving parts, simplicity and reliability also makes thermoacoustics attractive for isolated equipment.\cite{Ceperly1979} Examples of this are outer space applications\cite{Backhaus2004a,Tward2003,Garrett1993} and remote sensing techniques.\cite{Ali2013,Garrett2016a} Further advantages of using thermoacoustic devices are that they work on noble and inert gases (e.g. helium and argon) and that they do not rely on a phase change during the thermodynamic cycle.\cite{Avent2015b} Therefore, no harmful, ozone depleting refrigerants are needed for heat pumping purposes and operation over a wide range of temperatures is possible since the devices do not have to operate around a phase transition temperature.

Despite the promise of thermoacoustics, there is still plenty of work to do before the high theoretically possible efficiencies are approached in practice in a cost-effective manner. The low costs are especially important when utilizing low-grade heat, which is an area the authors think could be the most prominent application of thermoacoustics, since it can be most competitive in this area. As long as there is an abundant amount of low-grade (waste) heat available, minimizing the costs is of the main importance, even if at the expense of slightly reducing the efficiency. 

\subsection{Converting thermoacoustic power into electricity \label{sec:intro:thermoacousticselectricity}}
\begin{figure*}
	\centering
	\includegraphics[width=0.8\textwidth]{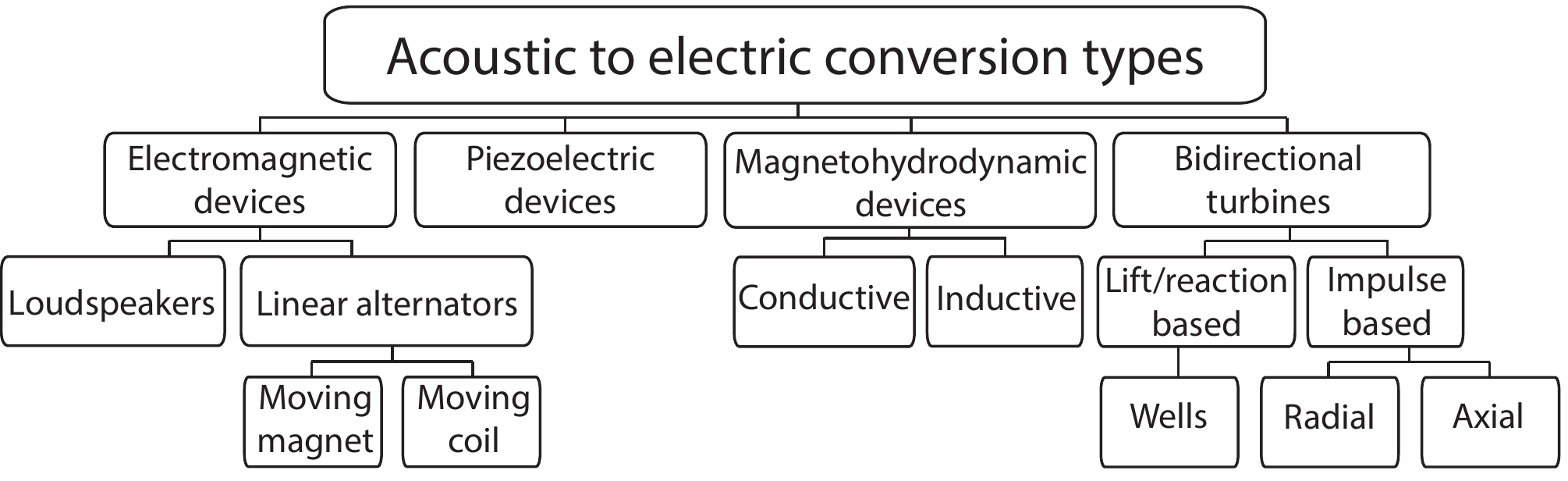}
	\caption{Overview of different methods for the acoustic to electric conversion.}
	\label{fig:Acoustic_to_electric_overview}
\end{figure*}

\noindent Despite the fact that most work in thermoacoustics has focused on refrigeration devices,\cite{Wollan2002,Keolian2015,Chen2000} producing electricity from acoustic power using a thermoacoustic engine also shows a high potential. As outlined in the previous section, wherever there is a sufficient heat source available, these engines can potentially produce electricity in a reliable and cost-effective way. \cite{Chen2013,Yu2010} So far, most thermoacoustic engines have relied upon electromagnetic components to convert acoustic power into electricity.  This was either by using a relatively cheap commercial loudspeaker in reverse or by making use of the more expensive but dedicated linear alternator. Other options that have been used in practice are piezoelectric components, magnetohydrodynamic devices, and bidirectional turbines. A schematic overview of these four options with some subdivisions is shown in Fig.~\ref{fig:Acoustic_to_electric_overview}.   

For people interested in producing electric power through the use of thermoacoustics, the authors feel the need for a clear overview of the different possibilities with accompanying advantages, disadvantages and recommended areas of application. This should attract designers to the area of thermoacoustics and ensure that they can more easily develop an engine suited for their needs. Besides the overview of current literature, knowledge on gaps and conflicts in current literature should be pointed out, which will make it easier to subsequently provide guidelines about where future thermoacoustic research should be headed. In part these requirements have been met by previous review papers, such as the ones by Avent\cite{Avent2015b} and Pillai.\cite{Pillai2014a} However, these articles generally provide a broader view on thermoacoustic energy harvesting, resulting in only a small amount of attention for the acoustic power to electricity conversion. Therefore, some of the conversion methods are not treated and/or not enough in depth knowledge of these is given. 

The authors provide an acoustic to electric specific review paper in this work, where all the topics pointed out in the previous paragraph are included. This will be done in individual sections for the four conversion methods that were previously mentioned and are shown in Fig.~\ref{fig:Acoustic_to_electric_overview}. Electromagnetic devices are treated in Sec.~\ref{sec:electromagnetic}, piezoelectric devices in Sec.~\ref{sec:piezoelectric}, magnetohydrodynamic devices in Sec.~\ref{sec:magnetohydrodynamic}, and bidirectional turbines in Sec.~\ref{sec:bidirectional_turbines}. After the individual methods, conclusions about these specific sections as well as general conclusions and recommendations by the authors are given in Sec.~\ref{sec:conclusions}.

\section{Electromagnetic devices \label{sec:electromagnetic}}
\noindent This section will provide a review of electromagnetic devices for converting thermoacoustic power into electricity. This will contain the devices that directly use electromagnetic induction to convert acoustic power into electrical power. Besides magnets and coils as the main components for the electromagnetic induction, iron is nearly always present as well for lower costs and a higher transduction efficiency. The main principle of the devices is that the acoustic power initiates mechanical movement of one component relative to the other two, therewith inducing an electric current in the coil. There are different configurations possible to ensure this relative movement. An overview of the options used in thermoacoustics will be given in Sec.~\ref{sec:electromagnetic:configurations}.

After introducing the different configurations, details about the use of them in thermoacoustic engines is given in Sec.~\ref{sec:eletromagnetic:characteristics}. This will include advantages, disadvantages and details of different aspects of these electromagnetic devices in the field of thermoacoustics. For the latter, the most important topics are the efficiency of the acoustic to electric conversion, the maximum electric power output, and the coupling between the acoustic field and the electromagnetic component.  

\subsection{Configurations \label{sec:electromagnetic:configurations}}
\begin{figure}
	\centering
	\includegraphics[width=0.4\textwidth]{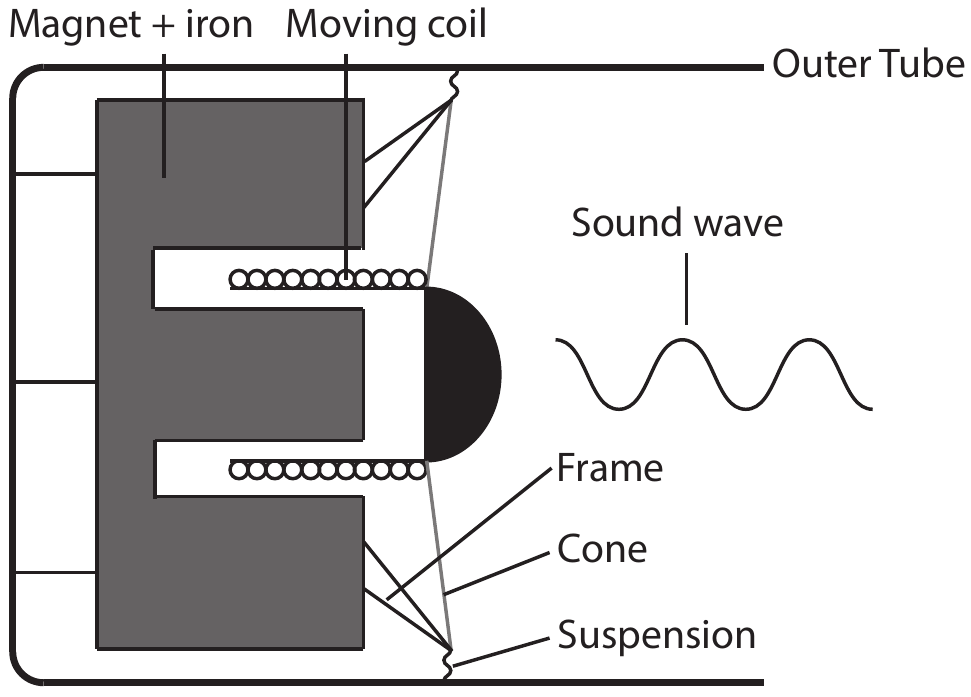}
	\caption{Schematic of a moving coil loudspeaker.}
	\label{fig:Loudspeaker_schematic}
\end{figure}

\noindent Conventional loudspeakers use electromagnetic induction to produce mechanical movement from electrical power. However, the loudspeakers can also be used in reverse to produce electricity, just as an electric motor can be used in reverse as a generator. Most loudspeakers use a moving coil in combination with static permanent magnets and iron, as shown schematically in Fig.~\ref{fig:Loudspeaker_schematic}. An incident sound wave will force the cone to move axially, therewith moving the connected coil with respect to the permanent magnets and inducing an electric current. 

Besides using a loudspeaker in reverse, one can also use the group of devices dedicated for converting acoustic power to electricity, namely linear alternators. In principle loudspeakers and linear alternators can be seen as the same electromagnetic devices. However, the authors feel a distinction is appropriate since the main purpose of loudspeakers is to produce sound with a flat response over a large frequency range. In contrast, linear alternators are designed to convert acoustic power to electricity at a single resonance frequency, which can result in quite different characteristics for both devices. This distinction can be seen in Fig.~\ref{fig:Acoustic_to_electric_overview}, where the linear alternator is further divided into moving magnet and moving coil devices, whilst practically all loudspeaker configurations use a moving coil to minimize the amount of moving mass.

\subsubsection{Loudspeakers}
\noindent A typical schematic representation of a moving coil loudspeaker in a tube is shown in Fig.~\ref{fig:Loudspeaker_schematic}. Such commercial loudspeakers are used in thermoacoustics mainly because they are relatively cheap and readily available. This makes them well suited for simple, initial experiments and low cost applications such as envisaged in the SCORE (Stove for Cooking, Refrigeration and Electricity supply) project. \cite{Jaworski2013a,Chen2013,Abdoulla-Latiwish2017a} However, loudspeakers usually have a poor power-transduction efficiency,\cite{swift2017thermoacoustics} since their design is more focused on linearity than efficiency \cite{Rossi2009a} and because they are designed to have a flat response over a large frequency range. Although loudspeakers should not generally be preferred over linear alternators, they can still be used if low costs are extremely important. This could be the case in situations where there is an abundant amount of (usually low-grade) heat. Note that loudspeakers are mainly used in low power thermoacoustics, due to their weak and fragile paper cones, their limited stroke lengths, and poor impedance match at a high mean pressure of the gas. \cite{swift2017thermoacoustics} Loudspeakers are not generally suitable at high power and acoustic amplitudes where there is a high pressure difference across the cone, but they may still be well usable when this pressure difference is in the kPa range. \cite{Jaworski2013a} Furthermore, robustness against more extreme operating conditions can be acquired by replacing the conventional cone with tougher yet still lightweight materials, such as aluminum or carbon fiber.\cite{Yu2012b} Tijani \textit{et al.} have shown this concept by modifying a loudspeaker with a \SI{0.1}{mm} thick aluminum cone.\cite{Tijani2002}

To select a suitable loudspeaker one can look at the procedure set out by Kang \textit{et al.},\cite{Kang2015a} which is based on the method by Yu \textit{et al.},\cite{Yu2011a} where both mainly focus on the acoustic coupling (see Sec.~\ref{sec:electromagnetic:characteristics:coupling} for more details) between the loudspeaker and the rest of the thermoacoustic engine. Besides these works, there are several other representative applications of loudspeakers in thermoacoustic engines. \cite{Yu2012b,Chen2013,Jaworski2013a} Typical operating pressures, efficiencies and power outputs for loudspeakers in thermoacoustic engines are given in Sec.~\ref{sec:electromagnetic:characteristics:efficiency}.

\subsubsection{Linear alternators}
\begin{figure}
	\centering
	\includegraphics[width=0.4\textwidth]{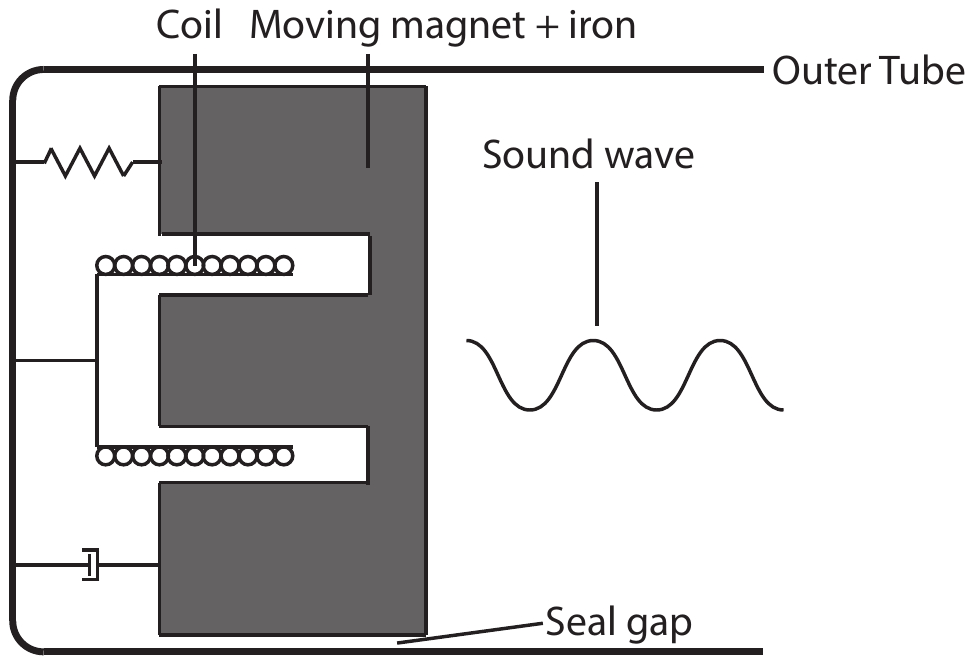}
	\caption{Schematic of a moving magnet linear alternator.}
	\label{fig:Linear_alternator_schematic}
\end{figure}

\noindent The term linear alternators is used for all devices that are dedicated for converting acoustic power into electricity using electromagnetic induction. In thermoacoustics there are two main configurations for linear alternators: moving coil and moving magnet. The moving coil designs use the same principle as the loudspeaker shown in Fig.~\ref{fig:Loudspeaker_schematic}, but don't have the weak and fragile paper cone that commercial loudspeakers have. For the moving magnet linear alternators the iron is either connected to the moving magnet, as is shown schematically in Fig.~\ref{fig:Linear_alternator_schematic}, or the iron remains fixed. In either case, the structure with the coil is fixed and the permanent magnet is forced to move axially (or linearly) by the incident sound wave. Although both moving coil and moving magnet alternators have been widely studied as a power producing component in thermoacoustics, the moving magnet linear alternators are more extensively used. It is worth noting that the linear alternators could also have a configuration where only the iron is moving, which is sometimes done in an effort to remove the costly permanent magnets.\cite{Rossi2009a} However, there is only a single example in the thermoacoustic literature of such an alternator.\cite{antonelli2017} Due to this lack of practical applications and available literature, the moving iron alternators are omitted in the rest of this work. A wide range of moving coil, magnet and iron linear alternator configurations for general purposes is presented in the book of Boldea and Nasar. \cite{Boldea1997}  

\begin{figure}
	\centering
	\includegraphics[width=0.5\textwidth]{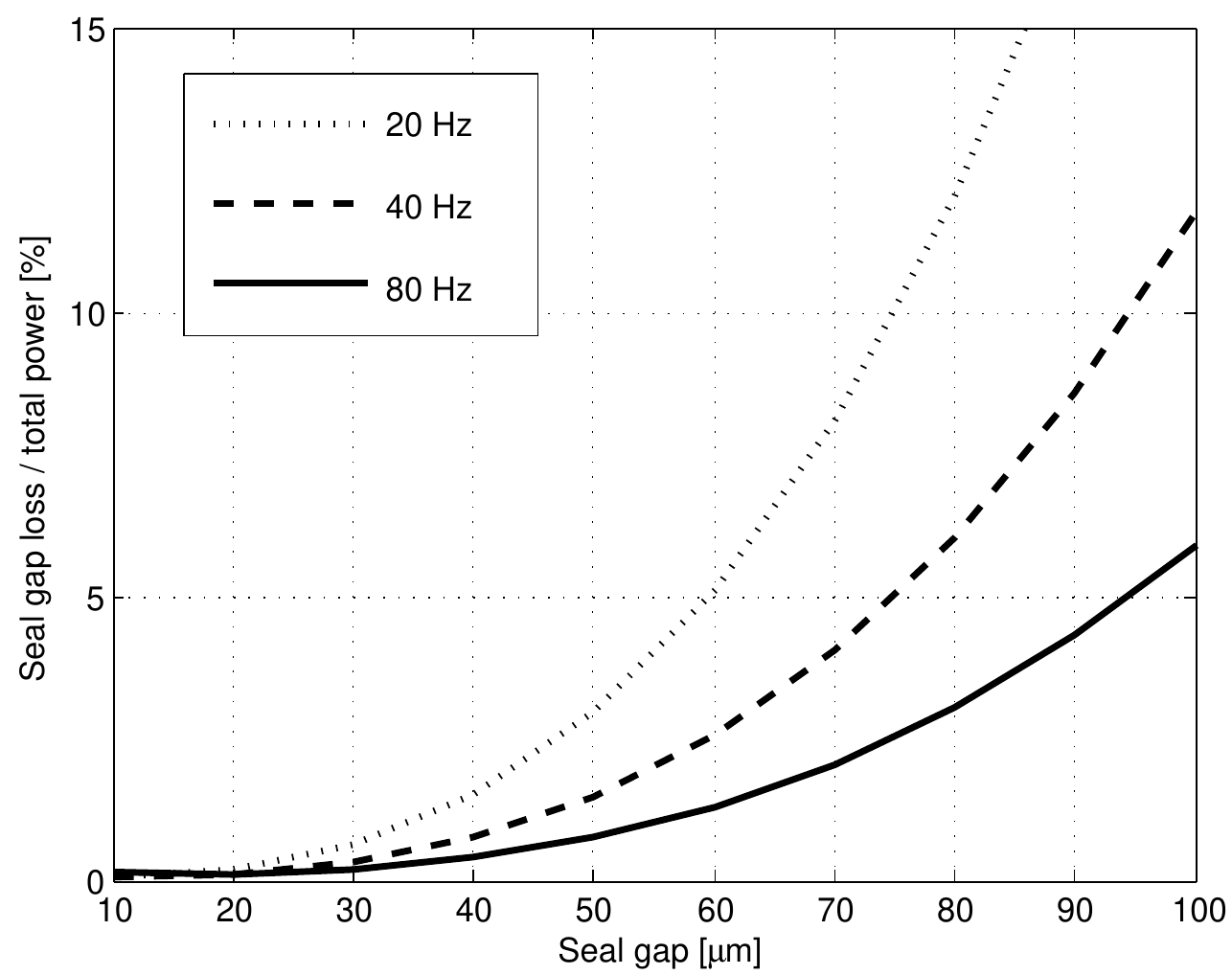}
	\caption{Calculated relative seal gap losses of a linear alternator as a function of the seal gap width for several frequencies.}
	\label{fig:seal_gap_losses}
\end{figure}

For any linear alternator design, flexure bearings ensure a very robust and reliable design. \cite{Backhaus2004a,Bailey2001} These metal plates provide a low stiffness in axial direction but a high resistance to radial and rotational movement. Therefore, the linear alternator can oscillate in the outer tube with seal gaps (see Fig.~\ref{fig:Linear_alternator_schematic}) as small as 10 \si{\mu m} without wearing against the outer tube. \cite{Backhaus2004a} These small gaps are essential to reduce blow-by and viscous friction inside this seal. \cite{Gonen2015b} Fig.~\ref{fig:seal_gap_losses} illustrates the magnitude of the seal gap losses with respect to the total alternator power as a function of the seal gap width. This represents a combination of basic calculations on the blow-by losses and shear losses of a typical, commercial linear alternator (John Corey, Qdrive, private communication, 2002). The results show that the seal gap losses increase severely for larger seal gaps, where the relative power loss can easily reach \SI{5}{\%} or more. To have an efficient linear alternator, one should therefore pay close attention to minimizing the seal gap dimensions. Furthermore, it can be seen from Fig.~\ref{fig:seal_gap_losses} that a higher acoustic frequency will cause relatively less power dissipation. This is caused by the linear increase of piston power as a function of frequency, whilst the seal gap losses only slightly increase as a function of frequency due to increasing shear losses.  

As an alternative to linear alternators with a seal gap, one could use flexure seal designs based on metal or composite bellows.\cite{garrett2004patent,poese2004patent} These designs do not require an axial alignment that must be accurate to within a few microns and eliminate the existence of the seal gap with accompanying losses. So far, these linear alternators have only been designed as acoustic drivers, where the seal gap is eliminated by bellows that connect the piston to the thermoacoustic refrigeration part in a flexible manner.\cite{garrett2004patent,poese2004patent} A similar design can be envisioned for the acoustic to electric conversion, where the bellows should connect the outer tube with the piston of the alternator, however these designs are yet to be constructed and tested.     

Thermoacoustic engine designs with linear alternators often use a double-acting configuration where the alternators are placed in pairs. \cite{Backhaus2004a,Wu2011a,Wang2016,Wang2017} If it is ensured that the alternators are phased correctly, they can counteract each other's vibrations. This provides a balanced design that is much less affected by spurious vibrations than a single acting configuration can have. This idea can of course be extended to more linear alternators, as long as they have a plane of symmetry such that they can balance each other.

Compared to commercial loudspeakers, linear alternators generally have a smaller range of mechanical resonance, a larger mass, and a higher power-transduction efficiency. \cite{swift2017thermoacoustics} They can also produce much more electrical power since they have larger stroke lengths and can operate at higher mean pressures and larger acoustic amplitudes. However, these dedicated components require more precision manufacturing, e.g. to minimize the losses through the seal gap shown in Fig.~\ref{fig:Linear_alternator_schematic}, and are made in much smaller volumes, therewith also making them significantly more expensive. Commercial linear alternators generally cost a few thousand dollars, while loudspeakers of similar dimensions cost around a hundred dollars. A linear alternator for large powers with a high efficiency can therefore easily be the most expensive part of a thermoacoustic engine. Furthermore, the mass of the linear alternator is the main contributor to the overall mass of a thermoacoustic engine. \cite{Backhaus2004a} Nevertheless, compared with loudspeakers the increased robustness, higher efficiency and larger power output of the linear alternator still make it a generally better option. An overview of linear alternator power outputs and efficiencies that have been achieved in thermoacoustics are given in Sec.~\ref{sec:electromagnetic:characteristics:efficiency}, where this can also be compared with that of commercial loudspeakers. 

\subsection{Characteristics \label{sec:eletromagnetic:characteristics}}
\noindent With the different types of electromagnetic devices described in the previous section, it is interesting to look at several performance characteristics such as efficiency and power output (see Sec.~\ref{sec:electromagnetic:characteristics:efficiency}) to be able to compare these devices. Furthermore, some other aspects such as the coupling of the electromagnetic device with the acoustic field (see Sec.~\ref{sec:electromagnetic:characteristics:coupling}) and analytical and numerical methods to study these devices (see Sec.~\ref{sec:electromagnetic:characteristics:other}) are also treated.

\subsubsection{Efficiency and electric power \label{sec:electromagnetic:characteristics:efficiency}}
\noindent An overview of the performance of several electromagnetic devices, often situated in an entire thermoacoustic engine, is given in Table~\ref{tab:electromagnetic}. The operating frequency for these typical thermoacoustic applications is in the range of \SI{50}{Hz} to \SI{150}{Hz}. The most used devices are moving magnet linear alternators, which are often homemade. Although the exact dimensions of the homemade designs are often not given, their listed mechanical and electrical properties such as amount of moving mass and the transduction coefficient can still be used to characterize the device. Furthermore, two commercial alternators are also used and listed in Table~\ref{tab:electromagnetic}, namely one by Lihan Thermoacoustic Technologies\cite{Wang2015} and one by Qdrive (Chart Industries).\cite{Wang2017} It is noted that the Qdrive 2s297 is originally designed as a compressor for cryocoolers\cite{Wang2017}, but still reaches an acoustic to electric conversion efficiency of \SI{73}{\%}. Dedicated linear alternators, if produced with high precision and used properly, can reach efficiencies in the range of \SI{80}{\%}-\SI{90}{\%}. These alternators will generally be quite expensive, which is why often an attempt is made to produce homemade alternators for prototype engines. As can be seen from Table~\ref{tab:electromagnetic}, these alternators still have an efficiency in the range of \SI{65}{\%}-\SI{75}{\%}. 

	\begin{table*}
		\caption{Overview compiled from literature of different types of electromagnetic devices with operating characteristics. Subsequent publications on the same work are grouped within lines. With $\eta_{a2e}$ the acoustic to electric conversion efficiency,  $\eta_{h2e}$ the efficiency from heat input to electricity, $P_{elec}$ the amount of electric power generated by the entire engine, $\Delta T$ the temperature difference of the heat supply of the engine, $P_{mean}$ the mean operating pressure, drive ratio the percentage of the pressure amplitude divided by the mean pressure, and device specs the specifics of the electromagnetic conversion device if available. All values are from experimental work, unless denoted with an asterisk ($^*$), and are the maximum values reported in the papers. A notion of n/a means the data could not be found in the paper and n/r means the field is not relevant since it does not apply for the given work.}	
		\centering
		\resizebox{\textwidth}{!}{
			\begin{tabular}{cccccccccc} 
				Type & Year$^{Ref}$ & $\eta_{a2e}$[\%]  & $\eta_{h2e}$[\%] & $P_{elec}$ [\SI{}{W}] & $\Delta T$ [\SI{}{K}] & $P_{mean}$ [bar] & Drive ratio [\%] & Method & Device specs \\ 
				\hline
				\rule{0pt}{2.5ex}    
				Moving coil alt. & 2004\cite{Backhaus2004a} & 75 & 18 & 39 & 620 & 55 & 9.8 & num/exp & n/a \\     
				\hline
				\rule{0pt}{2.5ex} 
				Moving magnet alt. & 2011\cite{Wu2011a} & 68 & 15 & 481 & 625 & 35 & 4.8 & exp &  homemade \\ 
				Moving magnet alt. & 2012\cite{Wu2012} & 65 & 15 & 481 & 625 & 35 & 5.0 & exp &  homemade \\ 
				Moving magnet alt. & 2014\cite{Wu2014a} & 74 & 20 & 1043 & 635 & 40 & 6.5 & num/exp & homemade \\ 
				\hline
				\rule{0pt}{2.5ex} 
				Moving magnet alt. & 2014\cite{Wu2014} & n/a & 17 & 1570 & 620 & 50 & n/a & num/exp & homemade \\ 
				Moving magnet alt. & 2015\cite{Bi2015} & 68 & 18 & 4690 & 625 & 60 & 7.5 & num/exp & homemade \\ 
				\hline
				\rule{0pt}{2.5ex} 
				Moving magnet alt. & 2013\cite{Sun2013a} & n/a & 12 & 345 & 620 & 30 & 4.0 & num/exp & homemade \\ 	
				\hline
				\rule{0pt}{2.5ex} 
				Moving magnet alt. & 2016\cite{Wang2016} & 51\cite{Wang2015} & 16 & 750 & 620 & 32 & 5.4 & num/exp & commercial\cite{Wang2015} \\ 	
				\hline
				\rule{0pt}{2.5ex} 
				Moving magnet alt. & 2017\cite{Wang2017} & 73 & n/a & 2300 & 390 & 40 & 9.3 & num/exp & Qdrive 2s297 \\ 	
				\hline
				\rule{0pt}{2.5ex} 
				Loudspeaker & 2011\cite{Yu2011a} & 60 & n/r & 3 & n/r & 1 & 1 & exp & B\&C 6PS38 \\ 
				\hline
				\rule{0pt}{2.5ex} 
				Loudspeaker & 2012\cite{Yu2012b,Yu2010} & 47 & 1 & 12 & n/a & 1 & 4.8 & num/exp & B\&C 6PS38 \\ 
				\hline
				\rule{0pt}{2.5ex} 
				Loudspeaker & 2013\cite{Chen2013} & 35 & n/a & 23 & 675 & 1.5 & n/a & num/exp & B\&C 6PS38 \\ 
				Altered loudspeaker  & 2012\cite{Saha2012} & 57 & n/r & n/a & n/r & 1 & n/a & num/exp & Halbach array \\ 
				Loudspeaker & 2017\cite{Abdoulla-Latiwish2017a} & 60 & 1.9 & 18 & 500 & 1 & 8 & num/exp & B\&C 8BG51 \\ 
				\hline
				\rule{0pt}{2.5ex} 
				Loudspeaker & 2015\cite{Kang2015a} & $45^*$ & 3 & 204 & 500 & 18 & $1.1^*$ & num/exp & B\&C 8NW51 \\ 			
			\end{tabular}
		} 
		\label{tab:electromagnetic}
	\end{table*}

Besides dedicated linear alternators, there has been quite some effort in using loudspeakers for the acoustic to electric conversion. As can be seen in Table~\ref{tab:electromagnetic}, the efficiency of the loudspeakers is in the range of \SI{35}{\%}-\SI{60}{\%}. It is interesting to see that the extremities of this range are reached with the same commercial loudspeaker (B\&C 6PS38), but in different experimental set-ups. Furthermore, the works using a loudspeaker focus on low-cost applications with a relatively small power output in the range of $\sim$$\SI{10}{W}$-\SI{200}{W}, where the reported \SI{200}{W} is actually for two loudspeakers.\cite{Kang2015a} For the linear alternators the highest power outputs in Table~\ref{tab:electromagnetic} are \SI{4690}{W} and \SI{2300}{W}, for six and two moving magnet linear alternators, respectively. This yields \SI{1150}{W} for the Qdrive linear alternator and \SI{100}{W} for a commercial loudspeaker, which is about an order of magnitude difference. It is still quite interesting that a standard commercial loudspeaker can be used at \SI{18}{bar} mean pressure with an amplitude of \SI{0.2}{bar} to produce \SI{100}{W} for at least a short period without rupturing.\cite{Kang2015a} However, much more power output and reliability can not be expected from loudspeakers. Therefore, one should either use a lot of loudspeakers under relaxed conditions, or preferably, use linear alternators if a power output larger than a few hundred Watt is desired. 

As shown in Table~\ref{tab:electromagnetic}, so far linear alternators in thermoacoustic engines have reached electrical power outputs in the \SI{}{kW} range. To increase this power to the \SI{}{MW} range, the dimensions of the thermoacoustic engine have to be increased, resulting in a lower operating frequency. This reduces the electromagnetic induction, causing the need for larger and stronger magnets which significantly increase the cost and mass of the alternator. Furthermore, a lower frequency means the alternator should have a larger stroke length within the small tolerance of the seal gap. These necessary adjustments result in a more than linear increase of alternator complexity and cost in terms of upscaling output power, which eventually constrains the practical and economical feasibility.         

\subsubsection{Coupling and impedance \label{sec:electromagnetic:characteristics:coupling}}
\noindent In thermoacoustic engines a working gas is used to transfer work in the form of acoustic power to an electricity generating component. Partly due to the large density difference between the gas and solid parts, this power transmission is not at all trivial. For a good coupling, the resonance frequency of the mechanical and electrical parts of the electromagnetic transducer should equal the working frequency of the engine.\cite{Sun2013a,AbdEl-Wahed2016} Furthermore, the transducer can be seen as an acoustic load. Therefore, acoustic impedance matching is also necessary for an efficient power transmission. Note that acoustic impedance is the ratio of the pressure amplitude over the flow rate, or in other words, the amount of driving force needed for a given volumetric displacement of the gas. 

Linear alternators generally need a large force for a small displacement, and can therefore be referred to as 'non-compliant' transducers and have to be placed in a high impedance region of the thermoacoustic engine. This high impedance generally leads to a high pressure drop,\cite{Yu2012b} therewith creating the necessity for a small seal gap (see Fig.~\ref{fig:Linear_alternator_schematic}) and precision engineering. Therefore, the high impedance and placement can ensure a high efficiency for alternators but also makes them relatively expensive. Loudspeakers are low impedance (small force large displacement) transducers and can be referred to as 'ultra-compliant'. They are generally placed in a low impedance region and therefore experience a smaller pressure drop. The fragile paper cone and limited stroke of loudspeakers limits them to produce a high power output. In an effort to improve this, one could take advantage of available loudspeaker technology and produce relatively cheap, robust and ultra-compliant alternators.\cite{Yu2012b}

Independent of the electromagnetic devices, a good acoustic coupling can be achieved by making sure the imaginary part of the acoustic load impedance is near zero whilst the real part is large. Wang \textit{et al.} use this to describe a procedure where they first analyze the impedance of the alternator and rest of the engine separately, after which they utilize this to optimize the acoustic coupling (by varying the alternator load resistance).\cite{Wang2016a} Other works have achieved similar results by varying the electric capacitance of the linear alternator\cite{Bi2015} and optimizing the position of it in the thermoacoustic engine.\cite{Kang2015a} It should be noted that the imaginary part of the acoustic load can also be used to tune the resonance circuit. Therefore, retaining a small imaginary part of the load can actually be beneficial for the acoustic coupling and therewith the engine performance. 

As shortly mentioned before, for an effective power transmission the resonance frequency of the electromagnetic device should equal that of the rest of the thermoacoustic engine. This can prove to be difficult, because placing such a device changes the acoustic field in the rest of the engine. Especially for linear alternators this can be a problem, since they only have a small frequency band in which they have resonance and therewith a high power transduction efficiency. The linear alternator dominates the coupling problem, resulting in the need to tune the acoustic circuit to the same resonance frequency. For loudspeakers, it has been shown that at typical operating frequencies of thermoacoustic engines the acoustic to electric efficiency can be relatively constant\cite{Yu2011a} (although lower than for linear alternators, as shown in Sec.~\ref{sec:electromagnetic:characteristics:efficiency}). Furthermore, the displacement amplitude only has a small influence on the efficiency, which is a good property for loudspeakers that depend on large stroke lengths due to their low impedance. \cite{Yu2011a} Nevertheless, for any device acoustic impedance coupling is still critical for a good power transduction efficiency. A few attempts to solve the coupling problem have been given in the previous paragraph. More options will be given in Sec.~\ref{sec:electromagnetic:characteristics:other}, where analytical and numerical methods to study electromagnetic devices and entire thermoacoustic engines are shown.  

\subsubsection{Other \label{sec:electromagnetic:characteristics:other}}
\noindent When developing or selecting an electromagnetic device, it is recommended to start this process with some analytical calculations to get an idea of the operating characteristics and performance. For this purpose, mathematical equations are often derived by using an electric circuit analogy for the alternators\cite{Wakeland2000,Sun2013a,Wang2016} and loudspeakers.\cite{Yu2011a} Yu \textit{et al.} validate experimentally that the acoustic to electric conversion efficiency can be accurately predicted in this manner.\cite{Yu2011a} In subsequent work they also use these calculations to select a commercial loudspeaker from fourteen possible options, whilst mainly focusing on the acoustic impedance for the highest power output and efficiency.\cite{Kang2015a} Gonen and Grossman extend the analytical calculations by including the velocity distribution, viscous friction, and blow-by losses in the seal gap to accurately predict the acoustic to electric conversion efficiency.\cite{Gonen2015a,Gonen2015b} Besides purely analytical calculations for the electromagnetic device, one can also use numerical simulations. Saha \textit{et al.} show the use of this by optimizing their double Halbach array linear alternator design by using 2D finite element method simulations.\cite{Saha2012} For a more elaborate mathematical background of calculations dedicated for linear electric generators one can look at the book of Boldea and Nasar.\cite{Boldea1997}    

To predict the performance of an entire thermoacoustic engine, one can add acoustic relations to the analytical calculations. The basis of the linear thermoacoustic theory was constructed by Rott\cite{Rott1980} and further developed by Swift.\cite{swift2017thermoacoustics} The set of equations following from this linear theory are often solved numerically by the Design Environment for Low-amplitude Thermoacoustic Energy Conversion (DeltaEC).\cite{DeltaEC} There are quite a few examples that show the successful use of DeltaEC for thermoacoustic engines with electromagnetic devices by giving the governing equations and experimentally validating the numerical results. \cite{Yu2012b,Bi2015,Wang2016} The examples show that for linear operating conditions (drive ratio $<\SI{10}{\%}$), DeltaEC can be a relatively accurate tool for predicting and optimizing the thermoacoustic engine performance before actually constructing and testing the engine.

Backhaus \textit{et al.} used a one-dimensional numerical model based on a lumped-element electric circuit analogous to the acoustic circuit.\cite{Backhaus2004a,Backhaus1999} Besides a good corresponence of the model with experiments, they also identify the different loss mechanisms of their linear alternator by carefully setting up different experiments.\cite{Backhaus2004a} This provides a nice overview of the magnitude of the alternator losses, which they use to optimize their device with a specific focus on reducing the total mass and volume for electricity generation aboard a spacecraft. The mass of the electromagnetic device is not only important in these exotic applications but, especially for the amount of moving mass, also in general. Attempting to increase the power output results in a higher moving mass, where eventually a practical limit is reached due to the difficulty to maintain a stable, large stroke amplitude in the seal gap under the extreme periodic forces.\cite{Blok2014}    

Since linear alternators can already achieve quite large power conversion efficiencies, it is important to focus on reducing the costs. This is in accordance with the rest of the thermoacoustic engine, which can be produced in a relatively cheap manner and therewith compete with alternative technologies. As of now, linear alternators are the most expensive component in thermoacoustic engines, which is why designers often choose for cheaper commercial loudspeakers at the cost of a lower efficiency and possible power output. To reduce the price of alternators, it is important to provide clarity about the costs that are made for a given work, especially if the linear alternators are homemade. In this manner, designers can learn which aspects are the most expensive and try to reduce these in future work. A good example, albeit for a thermoacoustic engine with a loudspeaker instead of linear alternator, can be found in the work of Chen \textit{et al.}\cite{Chen2013} They clearly show the costs of all components and the different stages in which they have reduced the costs of the thermoacoustic engine. Providing such information helps others to not only see how to reach a certain efficiency and power output, but also what the corresponding costs are per amount of power output.

\section{Piezoelectric devices \label{sec:piezoelectric}}
\noindent This section will provide a review of piezoelectric devices for converting thermoacoustic power into electricity. Piezoelectric materials produce electricity during mechanical deformation, which in the field of thermoacoustics is caused by the incident acoustic wave. Sec.~\ref{sec:piezoelectric:configurations} provides some piezoelectric materials that have been successfully used in thermoacoustic engines to produce electricity and gives information on the possible configurations for these devices. Subsequently, Sec.~\ref{sec:piezoelectric:characteristics} will focus on the characteristics of the piezoelectric components, such as power output, efficiency and coupling with a thermoacoustic engine. Furthermore, analytical and numerical methods to study thermoacoustic engines with piezoelectric materials are presented. Note that the focus of this section about piezoelectric devices is on the application in thermoacoustic engines. For the basic principles and mathematical background for piezoelectric devices one can look at general literature such as APC's book on piezoelectrics.\cite{piezoelectricAPC2002}

\subsection{Configurations \label{sec:piezoelectric:configurations}}
\noindent Piezoelectric materials, such as certain crystals and ceramics, build up a voltage difference across their opposite faces during mechanical deformation. If these sides are connected in an electrical circuit a current is induced, therewith producing electrical power. The reverse is also true, piezoelectric material will mechanically deform if it is connected in a circuit and a voltage difference is applied across it. The latter can be used in the field of thermoacoustics to produce acoustic waves, which can for example drive refrigerators. However, in this work we focus on the former, where the piezoelectric material is placed inside a thermoacoustic engine to convert the available acoustic power into electricity. An important side note is that one has to be careful with using piezoelectric material, because a large voltage potential can build up in an open electrical circuit, which can be very harmful if discharged upon human touch.     

The most widely used piezoelectric material for thermoacoustic purposes is the ceramic lead zirconate titanate, also referred to as PZT.\cite{wekin2008,Nouh2014b,Zhao2013} This is generally of a high quality and provides a good coupling,\cite{wekin2008} but its brittle nature also limits the strain it can experience without being damaged.\cite{Anton2007} Other materials used successfully in thermoacoustics include polyvinylidene fluoride, or PVDF, and piezoelectric fiber composite, or PFC.\cite{wekin2008} These materials can be used for their increased flexibility and resistance to cracking compared with PZT.\cite{Anton2007,Lee2004} Furthermore, one investigation has focused on the use of lead magnesium niobate-lead titanate (PMN-PT) crystals.\cite{Jensen2010a}  A more elaborate review on piezoelectric materials can be found in the work of Anton and Sodano.\cite{Anton2007} This includes the aforementioned materials amongst others, several tuning schemes, and different spatial configurations in which the material can be constructed and used.

The piezoelectric material is generally of a small mass with little inertia. It is therefore suitable for operating efficiently at a relatively high acoustic frequency and small wavelength.\cite{Smoker2012a,Avent2015b} This results in small resonators and therewith compact thermoacoustic engine designs. However, due to the small inertia the power output can also be limited, as will be shown in Sec.~\ref{sec:piezoelectric:characteristics:efficiency}. Lin \textit{et al.} have shown a weight can be added to increase the inertia and also tune the resonance frequency of the piezoelectric material.\cite{lin2016} Nouh \textit{et al.} also added a weight, but connected this with a spring to create a mass-spring system they refer to as a dynamic magnifier.\cite{Nouh2012,nouh2013thesis,Nouh2014c,Nouh2014} This component is placed between the acoustic resonator and the piezoelectric material in an attempt to enhance the strain experienced by the piezo-element for the same acoustic power. Note that they were inspired to use dynamic magnifiers by piezoelectric work from other fields of application.\cite{Cornwell2005,Ma2010,Aldraihem2011a}  

So far, all presented thermoacoustic work involving piezoelectric material has been for standing wave engine designs. Very little work is done on traveling wave engines with piezoelectric harvesters. Furthermore, the authors have not found any experimental traveling wave work, but only analytical and numerical research.\cite{Aldraihem2011,Nouh2013} The traveling wave work uses the classical Backhaus and Swift engine design,\cite{Backhaus1999,Backhaus2000} with the piezoelectric element placed at the end of the resonator. For the standing wave engines, the following configurations are identified: a straightforward tube section,\cite{Lihoreau2002a,Matveev2007} a Rijke tube based design,\cite{Zhao2013} a push-pull concept,\cite{Jensen2010a} the most commonly used Helmholtz-like resonators,\cite{Smoker2012a,nouh2013thesis,Nouh2014b,lin2016} and a looped-tube configuration with 'wagon wheel' style piezoelectric alternators.\cite{keolian2005patent,keolian2010thermacoustic,Keolian2011a} Note that the latter looped-tube designs by Keolian \textit{et al.} have traveling wave phasing in the regenerator but standing wave phasing at the piezoelectric element.\cite{Keolian2011a} Furthermore, these designs are quite different in comparison with the other designs because they have a large amount of piezoelectric elements and can produce significantly more power (see Sec.~\ref{sec:piezoelectric:characteristics:efficiency}).

\subsection{Characteristics \label{sec:piezoelectric:characteristics}}
\noindent This section will provide details on the performance characteristics of piezoelectric power harvesters in thermoacoustic engines. Achieved efficiencies and power outputs will be presented in Sec.~\ref{sec:piezoelectric:characteristics:efficiency}, followed by information on the impedance and coupling of the piezoelectric materials in Sec.~\ref{sec:piezoelectric:characteristics:coupling}. Furthermore, Sec.~\ref{sec:piezoelectric:characteristics:coupling} also outlines several analytical and numerical methods that have been used to investigate piezoelectric systems in thermoacoustics. Experimental validation of these methods is provided where it is available.

\subsubsection{Efficiency and electric power \label{sec:piezoelectric:characteristics:efficiency}}
\noindent An overview of achieved power outputs and efficiencies for piezoelectric harvesters in the field of thermoacoustics is given in Table~\ref{tab:piezoelectric}. Note that all values are for standing wave devices, since no output characteristics were found for the work on traveling wave piezoelectric engines.\cite{Aldraihem2011,Nouh2013} The most important point about the available research is that nearly all of it is for very low electric power output, namely in the order of a few milliwatt. This is an inherent result of using only one or a few small piezoelectric diaphragms. When used in this configuration, the thermoacoustic engines are not useful in producing significant amounts of energy from (low-grade) heat streams, and are therefore in a different field of application than the other energy harvesting methods reviewed in this paper. These amounts of electric power could however still be used in small and self powered, maintenance-free devices where physical access and electrical wiring is difficult, such as thermoacoustically powered remote sensors.

\begin{table*}
	\caption{Overview compiled from literature of piezoelectric harvesters in the field of thermoacoustics. With $P_{elec}$ the amount of electric power generated, $\eta_{a2e}$ the acoustic to electric conversion efficiency, the abbreviation of the used materials as outlined in Sec.~\ref{sec:piezoelectric:configurations}, and specifics of the piezoelectric material if available. All values are the maximum values reported in the papers. A notion of n/a means the data could not be found in the work and values denoted with an asterisk ($^{*}$) are normalized and non-dimensional and should therefore only be compared with each other.}
	\centering
	\resizebox{\textwidth}{!}{
		\begin{tabular}{ccccccc} 
			Authors & Year$^{Ref}$ & Method & $P_{elec}$ [\SI{}{mW}] & $\eta_{a2e}$[\%] &  Material & Piezoelectric specs  \\ 
			\hline
			\rule{0pt}{2.5ex}    
			Smoker \textit{et al.} & 2012\cite{Smoker2012a} & num/exp & 0.1 & 10 & PZT & Piezo Systems Inc. T107-A4E-573  \\     
			Zhao & 2013\cite{Zhao2013} & num/exp &  2 & 22 & PZT & Piezo Systems Inc. T216-A4NO-573X  \\  
			Wekin \textit{et al.} & 2008\cite{wekin2008} & exp &  0.2 & n/a & PZT,PVDF,PFC & only dimensions  \\  
			\hline
			\rule{0pt}{2.5ex} 
			Nouh \textit{et al.} & 2014\cite{Nouh2014c} & num/exp & 0.001  & 0.2 & PZT & normal, Digi-Key buzzer PZT-4  \\  
			Nouh \textit{et al.} & 2014\cite{Nouh2014c} & num/exp & 0.02  & 2 & PZT & dynamically magnified, Digi-Key buzzer PZT-4  \\   		
			\hline
			\rule{0pt}{2.5ex} 
			Jensen and Raspet & 2010\cite{Jensen2010a} & num & 400 & n/a & PMN-PT & characteristics given  \\  
			Matveev \textit{et al.} & 2007\cite{Matveev2007} & num & n/a & 10 & n/a & n/a  \\  		
			\hline
			\rule{0pt}{2.5ex} 
			Nouh \textit{et al.} & 2012\cite{Nouh2012} & num & 0.6 [-]$^*$ & 15 & n/a & normal, characteristics given  \\  
			Nouh \textit{et al.} & 2012\cite{Nouh2012} & num & 1.3 [-]$^*$ & 34 & n/a & dynamically magnified, characteristics given  \\  
			\hline
			\rule{0pt}{2.5ex} 
			Nouh \textit{et al.} & 2014\cite{Nouh2014} & num & 0.5 [-]$^*$ & 14 & n/a & normal, characteristics given  \\  
			Nouh \textit{et al.} & 2014\cite{Nouh2014} & num & 3 [-]$^*$ & 22 & n/a & dynamically magnified, characteristics given  \\  
			\hline
			\rule{0pt}{2.5ex} 
			Keolian \textit{et al.} & 2011\cite{keolian2005patent,keolian2010thermacoustic,Keolian2011a} & num & $4 \times 10^6$ & 90 & PZT & Morgan Electro Ceramics PZT807  \\  
			Keolian \textit{et al.} & 2011\cite{Keolian2011a,Keolian2011} & exp & $3.7 \times 10^4$ & n/a & PZT & Morgan Electro Ceramics PZT807  \\  
			
		\end{tabular}
	} 
	\label{tab:piezoelectric}
\end{table*}

The work by Jensen and Raspet approaches the field of significant power output with their simple thermoaocustic waste heat engine that produces \SI{0.4}{W} of electric power. However, this is based on numerical work only, and should therefore still be proven experimentally and subsequently scaled to an electrical output about two orders of magnitude higher to be of significant use. From all the results presented in Table~\ref{tab:piezoelectric}, the only work found in literature that is already in the order of significant power output is therefore that of Keolian \textit{et al.} They developed a thermoacoustic piezoelectric engine (and refrigerator) for a project where the heat produced by a heavy-duty diesel truck is used as an energy input. Details about their design can be found in their final report\cite{Keolian2011a} and patents.\cite{keolian2005patent,keolian2010thermacoustic} Numerical calculations have predicted their generator can produce \SI{4}{kW} of electrical power.\cite{Keolian2011a} So far, an early prototype has produced \SI{37}{W} and they claim that they can modify this prototype to produce \SI{600}{W}.\cite{Keolian2011,Keolian2011a} This is still less than predicted for their final design, however, it is significantly more electrical power than any other work has produced using piezoelectric harvesters in thermoacoustic engines. 

Keolian \textit{et al.} predict using an analytical calculation that their piezoelectric harvester can convert acoustic power into electricity with an efficiency of \SI{90}{\%} or higher.\cite{Keolian2011a,keolian2010thermacoustic} However, unfortunately the authors have not found any efficiency from their experimental work to confirm this calculation with such a high efficiency. As can be seen from Table~\ref{tab:piezoelectric}, the other works investigated show a maximum efficiency of around \SI{30}{\%} numerically and \SI{20}{\%} experimentally. These efficiencies are still quite low, since research from other fields show that it should be possible to use piezoelectric energy harvesters in the range of \SI{70}{\%} to \SI{90}{\%}.\cite{Yang2017}  

An improvement in the performance of piezoelectric devices has been made by Nouh \textit{et al.}\cite{Nouh2012,Nouh2014,Nouh2014c,nouh2013thesis} by introducing dynamic magnifiers, as explained in Sec.~\ref{sec:piezoelectric:configurations}. Table~\ref{tab:piezoelectric} lists the output power and efficiency for two numerical studies\cite{Nouh2012,Nouh2014} and one experimental work,\cite{Nouh2014c} where either a dynamic magnifier is used or not. The piezoelectric material and the rest of the thermoacoustic set-up is kept the same for this comparison. For both numerical works the efficiency approximately doubles by using a dynamic magnifier. Furthermore, the electric power is also increased by a factor of two and six. Besides this increase in performance, the critical temperature gradient for self-sustained oscillations is also significantly decreased by using a dynamic magnifier. For the experimental study, Nouh \textit{et al.}\cite{Nouh2014c} are aware that they use a piezoelectric material that is not as efficient as they have used in their previous experimental work by Smoker \textit{et al.}\cite{Smoker2012a} They still choose to work with this material since it is more practical for the purpose of choosing and optimizing several dynamic magnifiers. The efficiency and output power of their work is therefore quite low, as can be seen from Table~\ref{tab:piezoelectric}. Nevertheless, using their best dynamic magnifier they increase their power output by a factor 20 and their efficiency by a factor 10 compared with the case where no dynamic magnifier is used. Finally, it is worth mentioning that, although not as straightforwardly visible as in the work of Nouh \textit{et al.}, Keolian \textit{et al.}\cite{Keolian2011a,keolian2010thermacoustic} also use diaphragms and mechanical springs that could be classified as dynamic magnifiers.  

\subsubsection{Calculations and coupling \label{sec:piezoelectric:characteristics:coupling}}
\noindent To predict the performance of thermoacoustic engines with piezoelectric components, several analytical and numerical tools are available. Zhao has derived a nonlinear mathematical theory that shows a good agreement with experimental results on the acoustic velocity near the piezoelectric element.\cite{Zhao2013} However, his model does not include the performance characteristics of the piezoelectric element. Such governing equations are derived in several other works, where the coupling between the acoustic and piezoelectric part is done by impedance matching.\cite{Matveev2007,Jensen2010a,Nouh2014} These calculations give insight into the optimum efficiency and power output for a given engine design by varying the parameters of the piezoelectric transducer. Nouh \textit{et al.} search for this optimum by applying weights to several performance objectives and solving their equations numerically.\cite{Nouh2014} As described in the previous section, the performance characteristics from the aforementioned calculations can be found in Table~\ref{tab:piezoelectric}.

Aldraihem and Baz derive similar equations including the piezoelectric component, but subsequently use the root locus method from control theory to predict the onset of self-sustained oscillations for their traveling wave engine.\cite{Aldraihem2011} Nouh \textit{et al.} later use this same theory and compare it with an electrical analogy approach that combines the descriptions of the acoustic resonator and the stack with the characteristics of the piezoelectric diaphragm.\cite{Nouh2013,Nouh2014b} The analogous electrical circuit is numerically solved with the Simulation Program with Integrated Circuit Emphasis (SPICE) software.\cite{SPICE} They choose to use SPICE since it can be used for steady-state and transient analyses, in contrast with the widely used DeltaEC\cite{DeltaEC} that is limited to steady-state results. In their latest work, Nouh \textit{et al.} compare the frequency of oscillation for the root locus method, DeltaEC, SPICE and experimental results and show quite a good agreement.\cite{Nouh2014b} Furthermore, they show the tools can also predict the onset temperature gradient reasonably well. Note that the experimental power output and efficiency of the piezoelectric harvester for this specific thermoacoustic engine are presented in more detail in the work by Smoker \textit{et al.}\cite{Smoker2012a} Besides the acoustic part of the engine, this research shows that the predicted piezoelectric power output and efficiency of DeltaEC are also in very good agreement with that of their experiments.          

Inspired by the work of Smoker \textit{et al.},\cite{Smoker2012a} a CFD model of their complete standing wave thermoacoustic engine is build by Scalo \textit{et al.}\cite{scalo2015} and further developed by Lin \textit{et al.}\cite{lin2016} They solve the fully compressible Navier-Stokes equations in the entire engine, where the piezoelectric harvester is modeled as a multi-oscillator time-domain impedance boundary condition. This high-fidelity simulation shows a consistent match with the experimental work of Smoker \textit{et al.}, and provides an accurate tool for predicting the thermal-to-acoustic and acoustic-to-electric conversion efficiency.\cite{lin2016}  

The aforementioned methods, from simple analytical calculations to complex numerical simulations, can all be used to predict aspects of the thermoacoustic engine performance. An essential property of the piezoelectric energy harvesters is the conversion efficiency of acoustic power to electricity. To achieve a high efficiency, it is important to ensure a good coupling between the piezoelectric material and the acoustic resonator. This can be achieved by matching the acoustic impedance of the resonator with the mechanical impedance of the piezoelectric harvester.\cite{Aldraihem2011} Note that, as described in more detail in Sec.~\ref{sec:electromagnetic:characteristics:coupling}, acoustic impedance is the ratio of the pressure amplitude over the flow rate of the working gas. The mechanical impedance is similarly the force applied to the piezoelectric material over the induced velocity of the material. In practice, one can thus ensure a good coupling by altering the acoustic impedance through a change in operating conditions or dimensions of the thermoacoustic engine and/or by tuning the piezoelectric harvester. The latter can be done by e.g. varying the piezoelectric material and its configuration,\cite{wekin2008} adding a weight to the harvester to tune its resonance frequency,\cite{lin2016} or using a dynamic magnifier.\cite{Nouh2012,Nouh2014,Nouh2014c} 

\section{Magnetohydrodynamic devices \label{sec:magnetohydrodynamic}}
\noindent This section will provide a review of magnetohydrodynamic (MHD) devices for converting thermoacoustic power into electricity. The MHD transducers use an electrically conducting fluid that is forced to oscillate by the incident acoustic wave of the thermoacoustic engine. By applying a magnetic field over the oscillating working fluid, electricity can be produced in either an inductive or conductive manner. Since the working fluid is used to convert the acoustic power directly into electricity, there are no mechanically moving parts, just as in the rest of the thermoacoustic engine. Therefore, thermoacoustic engines with MHD generators are well suited for applications where no maintenance is possible, such as in outer space.\cite{Alemany2015,Alemany2014}

The working principle and details of both the inductive and conductive MHD transducers will be treated in Sec.~\ref{sec:magneto:configurations}, along with different working fluids that have been used in thermoacoustics. Subsequently, Sec.~\ref{sec:magneto:characteristics} will elaborate on the performance of the MHD generators and provide several methods for analyzing and optimizing this performance. This review is focused on the field of thermoacoustics, just as the review work of Hamann and Gerbeth from 1993.\cite{Hamann1993} For a more general introduction and mathematical background of magnetohydrodynamic devices one can look at general literature, such as the books by Rosa\cite{rosa1987} and Moreau.\cite{moreau1990} 

\subsection{Configurations \label{sec:magneto:configurations}}
\begin{figure}
	\centering
	\includegraphics[width=0.45\textwidth]{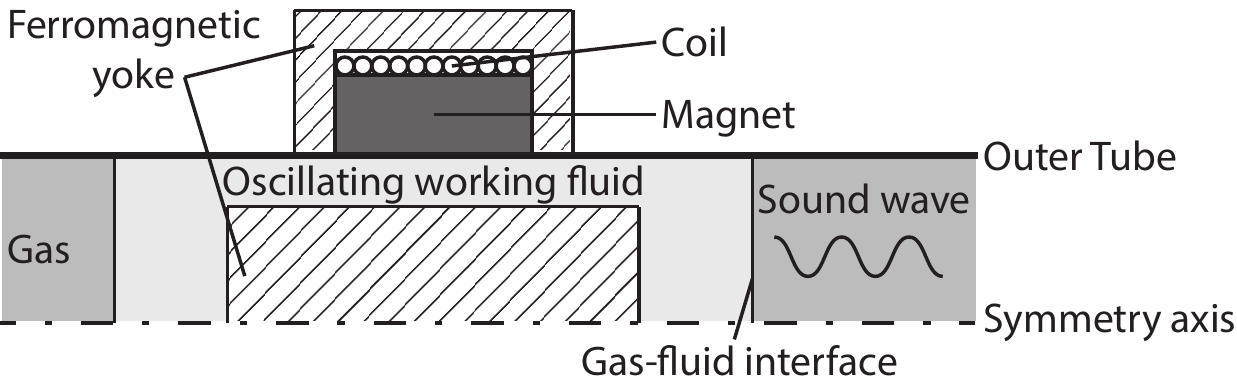}
	\caption{Schematic representation of an inductive magnetohydrodynamic transducer driven by an acoustic wave.}
	\label{fig:Magnetohydrodynamic_Inductive}
\end{figure}

\noindent The MHD generators can be divided in two main configurations: inductive and conductive. Fig.~\ref{fig:Magnetohydrodynamic_Inductive} schematically depicts an inductive MHD transducer. The permanent magnet supplies a magnetic field through the ferromagnetic yoke, with the magnetic field lines in the vertical direction of the schematic. The working fluid is driven by the sound wave to oscillate horizontally, and therewith perpendicular to the applied magnetic field. This will induce an alternating current in the fluid, which is accompanied by a pulsating magnetic field. The oscillatory motion of the fluid therewith induces an alternating current in the surrounding coil, which is placed in the direction perpendicular to both the permanent magnetic field and the oscillating fluid. In this manner, the inductive MHD generator creates an alternating electric current with adjustable strength and voltage.\cite{Mirhoseini2014a,Alemany2011a}     

The conductive MHD generators work similar to the inductive devices depicted in Fig.~\ref{fig:Magnetohydrodynamic_Inductive}. The main difference is that there is no coil, but a pair of electrodes that directly collects the alternating current from the working fluid. The electrode pair is placed perpendicular to both the magnetic field lines and the oscillating working fluid, which is in and out of the plane of Fig.~\ref{fig:Magnetohydrodynamic_Inductive}. The conductive MHD generators produce a strong electric current at a low voltage,\cite{Mirhoseini2014a,Alemany2011a} which is generally not preferred over the adjustable strength and voltage of the inductive MHD transducers.\cite{Vogin2007a} Furthermore, at high mean pressures of the working fluid, sealing the electrode connections from leaks might be troublesome.\cite{Alemany2011a} These disadvantages have caused the focus in thermoacoustics to shift from the conductive devices of the early days \cite{Swift1988b,wheatley1986,Hamann1993} towards inductive MHD generators in most of the recent work.\cite{Mirhoseini2014a,Alemany2011a,Alemany2014}   

Independent of the configuration, the electrically conducting working fluid mostly used is liquid sodium.\cite{Swift1988b,Alemany2015,Vogin2007a} This is a suitable working fluid due to its relatively low density and high electrical conductivity, which makes it sensitive to the influence of a magnetic field.\cite{Alemany2011a} Similarly, a saturated sodium bicarbonate aqueous solution is experimentally shown to be suitable as well.\cite{Castrejon2006a} Although it shows a good performance, one has to be careful with sodium since it is a highly reactive and possibly hazardous material. Swift \textit{et al.} have reported serious damage to their thermoacoustic engine due to a violent chemical reaction.\cite{Swift1988b} Besides sodium, any other electrically conducting fluid can be used, such as other liquid metals, salts and even plasma. For plasma, one can use a gas just as in the rest of a conventional thermoacoustic engine, and locally induce a plasma by periodically discharging a high voltage by electrodes immersed in the gas.\cite{Alemany2014}  

Swift \textit{et al.} have developed one of the first thermoacoustic engine prototypes with an MHD generator, as presented in a series of work\cite{Migliori1988,Swift1986,Swift1988b} and a patent application.\cite{wheatley1986} Their engine, including the prime mover generating the acoustic power, is completely filled with liquid sodium at a pressure of \SI{200}{bar}. Although the engine has shown quite a good performance (see Sec.~\ref{sec:magneto:characteristics}), the high pressure in the entire engine can cause problems in sealing the electrodes\cite{Alemany2011a} and generating the acoustic oscillations.\cite{Castrejon2006a} As an alternative, one could use a gas for the acoustic part of the engine and an electrically conducting fluid for the MHD generator.\cite{Alemany2011a,Alemany2015,DeBlok2014a} Separating the two fluids can for example be done by gravity using a U-shaped tube.\cite{Castrejon2006a} A downside of such designs is that there is a gas-fluid interface, as shown in Fig.~\ref{fig:Magnetohydrodynamic_Inductive}. Transmitting all of the acoustic power across the interface can be hard due to the large impedance difference\cite{DeBlok2014a} and the possibility of a Rayleigh-Taylor instability,\cite{Sharp1984} which can occur due to the gas with a smaller density periodically forcing the heavier fluid. For oscillating flows of sufficient amplitude, this instability of the gas-fluid interface has been observed in a magnetohydrodynamic fluid pump.\cite{Poese1996} The operating frequency of the two-fluid thermoacoustic engines is mostly around \SI{100}{Hz} or lower,\cite{Alemany2011a,DeBlok2014a} while that of the fully liquid sodium engines is in the order of \SI{1000}{Hz}.\cite{Migliori1988,Swift1988b}         

Since the design of the MHD generators is symmetric, it can be driven from either side. To increase the power output and reduce the necessary drive ratio in the acoustic part, several engine designs have been proposed that use this symmetry to drive the MHD generator from both sides in a so called push-pull mode.\cite{wheatley1986,DeBlok2014a,Alemany2011a,Alemany2015} As will be treated in more detail in the following section, these designs can improve the performance and relax the operating conditions of the thermoacoustic engine, but also result in a more complex coupling of the acoustic gas with the MHD working fluid.

\subsection{Characteristics \label{sec:magneto:characteristics}}
\noindent An important operating characteristic for MHD devices is the magnetic Reynolds number, where the kinematic viscosity of the ordinary Reynolds number is replaced with the magnetic diffusivity.\cite{Swift1988b} The magnetic Reynolds number is therefore a measure of the advection (or induction) of a magnetic field relative to the magnetic diffusion. In a series of work by Alemany \textit{et al.}, they do calculations for low magnetic Reynolds numbers, which allows them to neglect the induced field compared to the applied one for the MHD part.\cite{Alemany2011a} For the thermoacoustic part, they use a linearized version of the Navier-Stokes equations that is based on the work of Swift \textit{et al.}\cite{Swift1985} The resulting set of equations have provided them with a tool to analyze the performance of thermoacoustic engines with both conductive\cite{Vogin2007a} and inductive\cite{Alemany2011a,Mirhoseini2014a} MHD generators. For the conductive MHD generator they predict a maximum acoustic to electric conversion efficiency of \SI{80}{\%},\cite{Vogin2007a} although they later state that this was not for realistic load conditions.\cite{Alemany2011a} Subsequent calculations for an inductive MHD generator show a theoretical efficiency of \SI{65}{\%} and an electric power output of \SI{500}{W}.\cite{Alemany2011a} Further work on the inductive generator focuses on the relation between the acquired efficiency and the magnetic Reynolds number.\cite{Mirhoseini2014a} Through the working principle of an MHD generator, they explain the influence of the magnetic Reynolds number and show efficiencies in the range of \SI{60}{\%} to \SI{72}{\%}.    

The previous paragraph has given promising performance calculations for MHD generators in thermoacoustics, but unfortunately there is not much experimental work to confirm these calculations. One paper presents an experimental prototype of a conductive MHD generator that only produces a few millivolts.\cite{Castrejon2006a} No load is applied to measure the power output due to this small voltage, but a theoretical prediction based on the book by Rosa\cite{rosa1987} does show a good correspondence with the voltage output. The only experimental prototype with significant power in thermoacoustics is found in a series of work by Swift \textit{et al.}\cite{Migliori1988,Swift1986,Swift1988b,wheatley1986} As described in Sec.~\ref{sec:magneto:configurations}, they developed a thermoacoustic engine with a conductive MHD generator that is completely filled with liquid sodium at \SI{200}{bar}. The highest acoustic to electric conversion efficiency they achieve in their experiments is \SI{45}{\%} and they produce \SI{300}{W} of electric power.\cite{Swift1988b} They compare their results with calculations based on the linearized Navier-Stokes and Maxwell equations, and show a good correspondence for the conversion efficiency, with the experimental values being even slightly higher than their calculations predict. This good correspondence strengthens their claim that, with modest changes of their design that are based on analytical calculations, they expect to reach an efficiency of around \SI{70}{\%}.\cite{Swift1988b} Inspired by the work of Swift \textit{et al.}, an analytical optimization study was performed for an MHD generator with similar physical conditions.\cite{Ibanez2002} Characterized by four dimensionless parameters, the isotropic electrical efficiency and the overall second law efficiency, based on the entropy generation rate, were optimized for maximum performance. Besides the use for this single case, the entropy generation rate can be a useful tool for modeling and optimizing MHD generators in general.\cite{Ibanez2002}  

For an optimum performance of an MHD generator in a thermoacoustic engine, it is important to make sure there is a good coupling with the acoustic circuit. In the case of a combined gas-fluid engine, one has to ensure a close match of the natural frequency of the acoustic prime mover with the liquid MHD column for a good power transmission.\cite{Castrejon2006a} However, this can be hard due to the large density, and therewith impedance difference between the gas and liquid. De Blok \textit{et al.} use the impedance difference to calculate that for a typical engine and operating conditions \SI{90}{\%} of the acoustic power is expected to be reflected back at the gas-fluid interface.\cite{DeBlok2014a} To transfer significant acoustic power to the high impedance region of the MHD generator, their thermoacoustic drive ratio has to be \SI{15}{\%}, which will result in serious acoustic losses. Therefore, they propose to drive the MHD generator from two sides in a push-pull configuration to reduce the drive ratio (i.e. required pressure amplitude) by a factor 2. Alemany \textit{et al.} also use a push-pull configuration and use the iterative Newton's method to find the optimal conditions of their engine by minimizing the calculated impedance difference.\cite{Alemany2011a} The latter can result in a good acoustic power transmission, but one has to ensure that there remains a 180 degree phase difference between the two thermoacoustic parts driving the MHD generator. This can be a problem since the two prime movers tend to oscillate in phase.\cite{Spoor2000} De Blok \textit{et al.} explain how to acoustically ensure that the phase difference remains intact and propose a two-stage thermoacoustic engine design that numerically shows a good performance.\cite{DeBlok2014a} 

\section{Bidirectional turbines \label{sec:bidirectional_turbines}}
\noindent A bidirectional turbine is a special type of turbine that rotates in the same direction, independent of the direction of the axial flow. It is therefore suited to convert acoustic waves into rotational work, and subsequently into electricity when connected to a generator. Due to the relatively new introduction of bidirectional turbines in the field of thermoacoustics, the available literature is scarce. Nevertheless, the authors believe bidirectional turbines have potential because they are relatively cheap, can be coupled with highly efficient off-the-shelf generators and are promising for scaling to industrial sizes. Therefore, bidirectional turbines are still included in this review, but this section will be somewhat different than the others. A substantial amount of information will be given from the field of Oscillating Water Columns (OWC), where bidirectional turbines are used to convert marine wave energy into electricity. The focus will be on information from OWC literature that can be used to successfully implement bidirectional turbines in the field of thermoacoustics. In Sec.~\ref{sec:bidirectional_turbines:characteristics}, several bidirectional turbine designs with their most important features will be introduced, along with their advantages and disadvantages. Sec.~\ref{sec:bidirectional_turbines:characteristics} will start with performance results from the field of thermoacoustics, after which some important characteristics from the field of OWC's are given.  

\subsection{Configurations \label{sec:bidirectional_turbines:configurations}}
\begin{figure}
	\centering
	\includegraphics[width=0.48\textwidth]{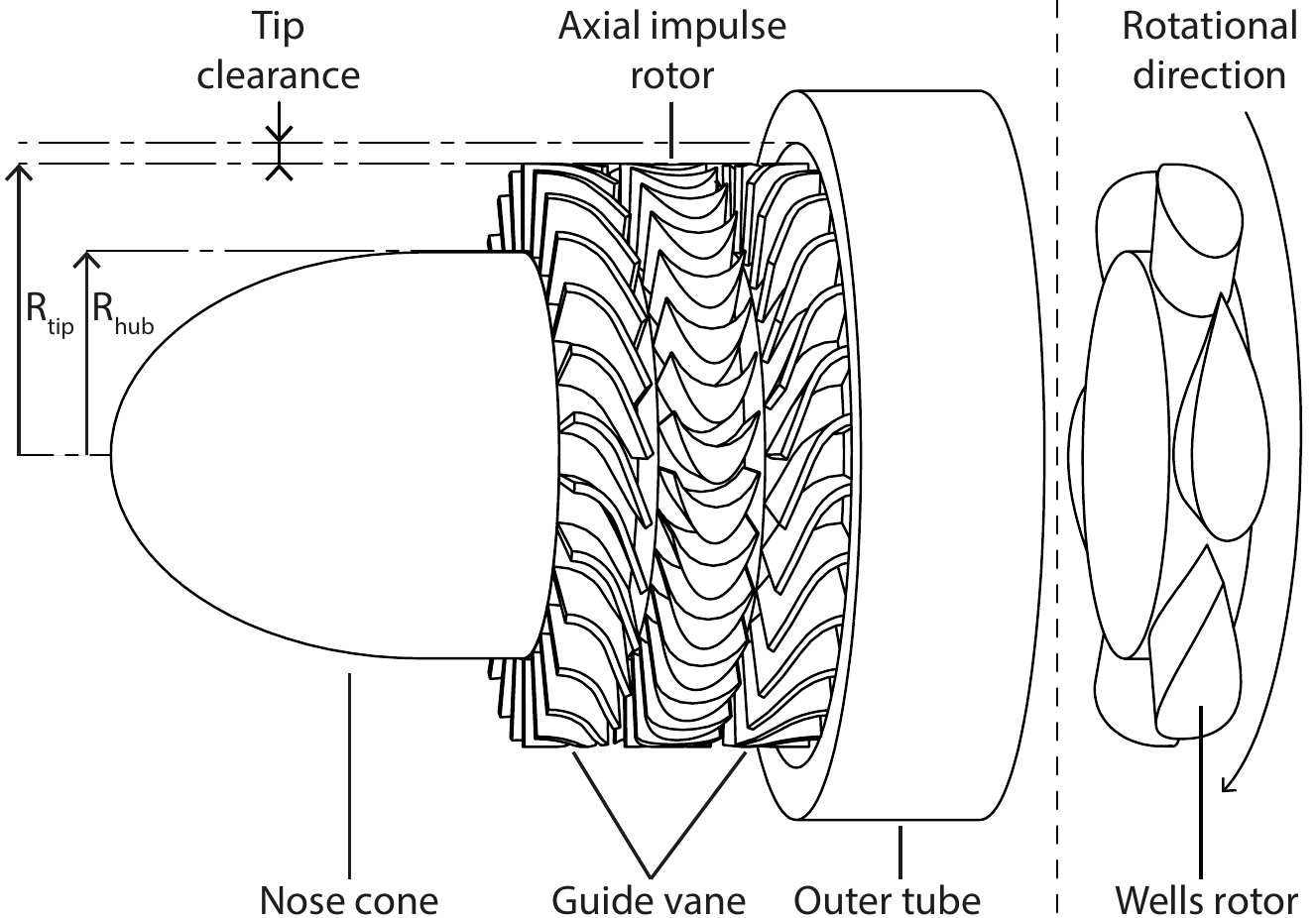}
	\caption{Schematic representation of an axial impulse turbine and the rotor blade of a Wells turbine, shown on the left and right hand side of the dashed line, respectively.}
	\label{fig:turbines}
\end{figure}

\noindent Bidirectional turbines, which can also be referred to as self-rectifying turbines, can be split into impulse and reaction based configurations. Fig.~\ref{fig:turbines} depicts an axial impulse turbine design, where the two static guide vanes convert the acoustic pressure energy to kinetic energy and direct this flow to the rotor blades. Due to the symmetric design, the rotor will experience a torque in the same rotational direction, independent of the incoming axial flow. The mechanical work of the rotor can then be converted to electricity by connecting it with a generator that can be placed inside the nose cone. This nose cone is generally used to guide the incoming flow towards the blades, where the latter do not span the whole radius but run from the central part called the hub towards the tip (see R$_{\textrm{hub}}$ and R$_{\textrm{tip}}$ in Fig.~\ref{fig:turbines}). Note that the guide vanes are usually kept in place with a ring around the blades that is connected to the outer tube, but this is not depicted in Fig.~\ref{fig:turbines} to show the shape of the blades. 

A similar, but less widely used configuration is the radial impulse turbine. The working principle of this turbine is the same as the axial version, but here the rotor is placed between the two guide vanes in radial direction, where one guide vane directs the flow outward towards increasing radius and the other inward towards the central axis. The kinetic energy of the bidirectional flow is again converted to rotational work by a similarly shaped impulse rotor. An advantage of the radial impulse turbine over the axial version is the reduced amount of axial thrust, which relaxes the fatigue on the bearings of the generator.\cite{Das2017} Besides this difference, both impulse turbines are characterized by a wide operating range with efficient performance and good starting characteristics, where the latter is ensured by the use of guide vanes. However, the large incidence angle of the flow at the downstream guide vane also causes an increase in aerodynamic loss, which can not be avoided due to the symmetry needed for operating in bidirectional flows.\cite{Karthikeyan2013}  

In contrast with the impulse turbines that use the velocity component, reaction based turbines use the pressure oscillations of the acoustic work to produce lift and therewith torque on a rotor blade. The most widely used reaction based turbine is the Wells turbine, of which the rotor blades have the shape of symmetrical airfoils, as can be seen at the right hand side of Fig.~\ref{fig:turbines}. The shape of the airfoil will cause a component of lift in the direction from the thin trailing edge to the blunt leading edge, therewith producing a torque in the rotational direction. Several airfoil shapes have been shown to work for bidirectional flows, including the most widely used NACA0015 through NACA0025 shapes.\cite{Boessneck2016,Setoguchi2006b} The Wells turbine can be used in combination with guide vanes, but these are often omitted since they are not necessary for the turbine to work (in contrast with the impulse turbines). Furthermore, the Wells turbine generally has more simple and fewer blades than the impulse turbines. These points make the Wells turbine the most simple and economic bidirectional turbine to convert acoustic power into electricity.\cite{Takao2012} However, there are several downsides during operation of the Wells turbine, such as the poor starting characteristics and the narrow range of efficient operation,\cite{Setoguchi2001,Takao2012} as will be discussed in more detail in Sec.~\ref{sec:bidirectional_turbines:characteristics}.   

Next to the distinction in turbine configuration, several design aspects are also important for bidirectional turbines in thermoacoustic engines. A wide range of nose cone designs is available for conventional turbines, of which hemispherical, parabolic and elliptical shapes have been used successfully in thermoacoustics.\cite{Boessneck2016} Another important aspect is the radial spacing between the edge of the rotor blades and the outer tube, called the tip clearance (see Fig.~\ref{fig:turbines}). Minimizing this distance is very important to ensure that the acoustic power is transferred to the turbine blades and does not leak along the outer edge. This importance is illustrated in bidirectional flows for both impulse\cite{Thakker2004b} and Wells\cite{Kim2002} turbines. In case guide vanes are used, another important spacing is the axial distance between the rotor and guide vanes, which should be minimized to reduce the leakage in radial direction and therewith increase the efficiency. Furthermore, the ratio of the number of guide vane blades over rotor blades is important. For an efficient energy transfer and no spurious vibrations, there should be a minimal amount of symmetry planes between the guide vane and rotor. This means that if the rotor has an even number of blades (e.g. for a good balance), then it is optimal to choose a prime (or coprime) number of blades for the guide vanes. Finally, note that besides the important aspects mentioned in this paragraph, there are several other design features that can be important for the turbine performance, such as blade solidity and the hub-to-tip ratio. An overview of the optimum design values for a wide range of aspects is given in the work of Das \textit{et al.} for Wells and impulse turbines in OWC's.\cite{Das2017}    

Just as in conventional steady flow turbines, the Wells and impulse turbines can be extended with additional rotors and guide vanes to introduce multiple stages.\cite{Setoguchi2006b} This can be done in an effort to enhance the coupling between the turbine and working gas and to increase the power output. For the guide vanes, an option used for OWC's is (self-)pitching guide vanes that pivot around a point to increase the efficiency of the turbine.\cite{Setoguchi2001,Karthikeyan2013} However, due to the relatively high acoustic frequency and the low-maintenance requirements for the rest of the thermoacoustic engine, it is encouraged to stick with fixed guide vanes for simplicity and reliability. With these fixed designs and the low material requirements due to the ambient temperature, the turbines can be easily manufactured with the cheap process of 3D rapid prototyping. Besides initial hardware costs that are currently dropping rapidly, the material costs of 3D printing a turbine is in the order of only ten dollars per \SI{}{dm^3} of material. When combined with the mature technology of conventional rotary generators, where the costs are approximately a hundred dollars per \SI{}{kW} of electric power output, the acoustic to electric conversion using bidirectional turbines can be relatively cheap and results in a reliable design.  

\subsection{Characteristics \label{sec:bidirectional_turbines:characteristics}}
\begin{figure}
	\centering
	\includegraphics[width=0.45\textwidth]{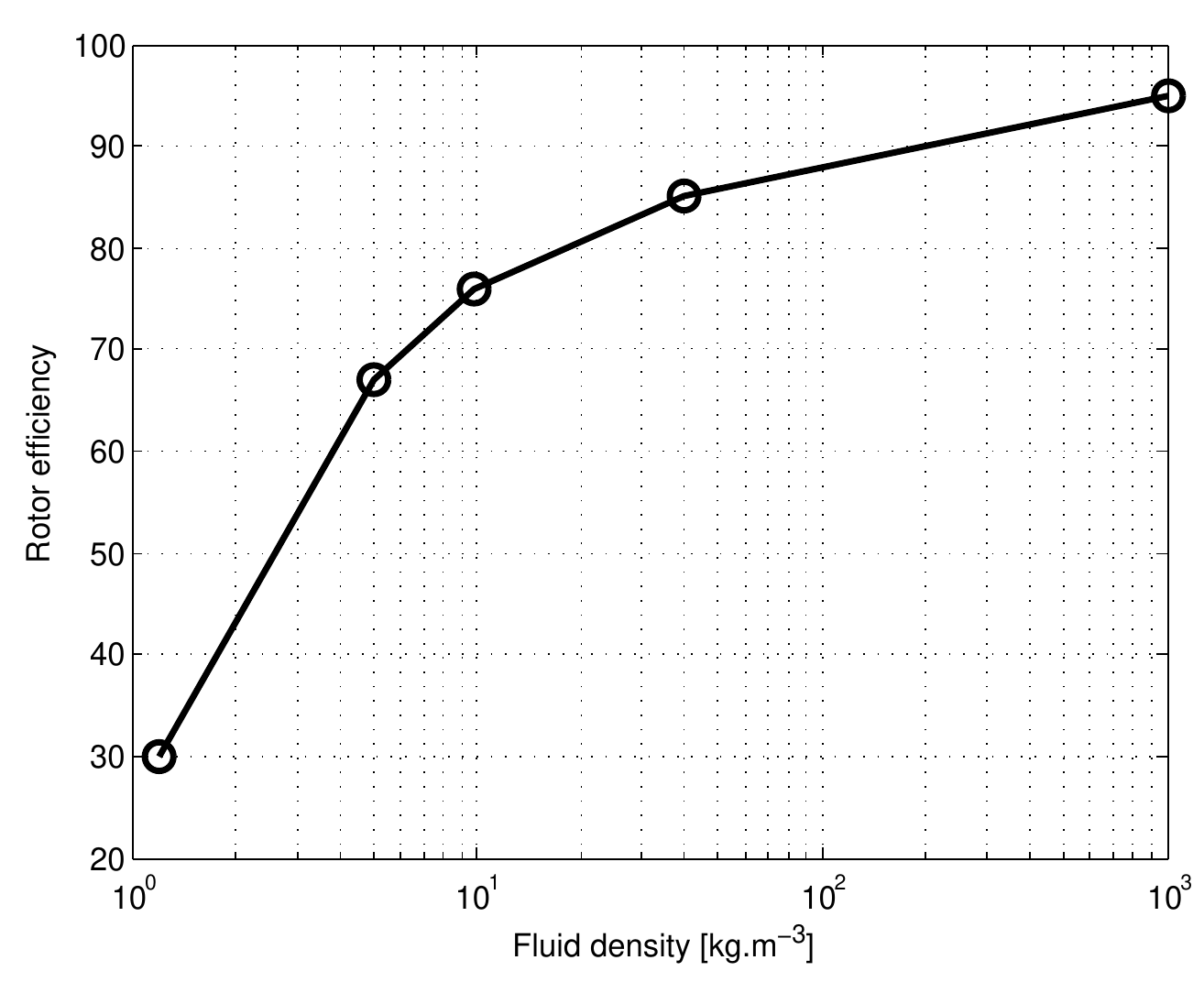}
	\caption{Bidirectional impulse turbine efficiency for various fluid densities as experimentally measured by de Blok \textit{et al.}\cite{Blok2014}}
	\label{fig:turbine_density_efficiency}
\end{figure}

\noindent The use of bidirectional turbines in thermoacoustic engines is fairly new compared with the other conversion technologies, as shown by the fact that the first published research is from 2014.\cite{Blok2014} In this work, de Blok \textit{et al.} experimentally investigate the use of a 3D printed bidirectional impulse turbine with varying acoustic power up to \SI{30}{W}. Besides the amplitude, they also varied the acoustic frequency in the range of \SI{20}{HZ} to \SI{50}{HZ}. Interestingly, they found no significant influence of the operating frequency on the turbine efficiency, suggesting that it is unnecessary to match the frequency of the acoustic circuit with the turbine for efficient power transduction in the engine. For air at ambient pressure, they measured the acoustic to electric efficiency to range between \SI{25}{\%} and \SI{30}{\%} for varying acoustic power. These values are in line with the performance acquired for impulse turbines in OWC's that operate at ambient pressure.\cite{Das2017,Falcao2016} However, thermoacoustic engines generally work at higher mean pressures, and thus larger mean fluid densities, since these are proportional for ideal gases. As measured by de Blok \textit{et al.},\cite{Blok2014} a larger mean density will significantly increase the efficiency of the impulse turbine, as shown in Fig.~\ref{fig:turbine_density_efficiency}. With a typical thermoacoustic engine operating at \SI{40}{bar} mean pressure (with density for air $\sim$\SI{48}{kg.m^{-3}}), they show that a rotor efficiency of \SI{85}{\%} can be acquired. When coupled with commercial generators that are highly efficient, converting acoustic power to electricity at an efficiency of \SI{80}{\%} is thus feasible. Furthermore, de Blok \textit{et al.} have confirmed these results for larger bidirectional turbines and see no limitations in scaling their systems to power levels in the \SI{}{MW} range.\cite{Blok2014} This confidence seems reasonable since generators and turbines generally become more efficient at increasing size.     

At present, two other works are available that investigate the use of bidirectional turbines for thermoacoustic engines. Kaneuchi and Nishimura have presented an extended abstract about an impulse turbine similar to that by de Blok \textit{et al.}, and find a rotor efficiency of around \SI{20}{\%}, which results in \SI{12}{\%} acoustic to electric conversion due to their inefficient generator.\cite{Kaneuchi2015} Boesnneck and Salem have investigated both Wells and axial impulse turbines with a wide variety of design changes, including different guide vanes and nose cones.\cite{Boessneck2016} Although they envision using their designs in thermoacoustic engines, unfortunately they have only performed steady flow measurements for their current research, which makes their investigation similar to OWC literature. Nevertheless, the extensive amount of designs have still resulted in some interesting bidirectional turbine characteristics, of which the most relevant are given next. For the Wells turbines they have done measurements with the NACA 0015, 0018 and 0021 rotor blades, and found no significant difference between the three. The optimum number of rotor blades is identified as five, and they find mixed results when using guide vanes for their Wells turbine, with a slight increase in performance possible when compared to the case without guide vanes. For the axial impulse turbine, they find that for their guide vanes a 12$^\circ$ inclination angle works better than a 30$^\circ$ one. Varying the blade angle for the rotor blades between 50$^\circ$  and 70$^\circ$ has not resulted in any significant performance change. The best nose cone is of an elliptical shape, and the rapid prototyping technique that produces the smoothest surface has resulted in the best performance. Finally, when comparing the Wells turbine with the axial impulse turbine, they conclude that the Wells turbine has a small range with superior performance, but the impulse turbine has much better starting characteristics and shows a more stable performance over a wide range of operating conditions.\cite{Boessneck2016}    

In the field of OWC's, the poorer starting characteristics of the Wells turbine have also been widely shown.\cite{Raghunathan1995,Setoguchi2001,Takao2012,Das2017} Besides the starting of the turbine at small flow rates, the angle of stall also bounds the operating range of the Wells turbine for high flow rates.\cite{Raghunathan1995} Furthermore, they generally run at a high rotational speed, which is accompanied by unwanted noise and a high axial thrust.\cite{Setoguchi2001,Karthikeyan2013} The impulse turbines have good starting characteristics and no problems of stall at higher flow rates, and therefore have the advantage of a wider operating range where a change in Reynolds number does not have much influence on the turbine efficiency.\cite{Das2017} However, this does come at the cost of a smaller peak efficiency then the Wells turbines, with the radial impulse turbine generally having a smaller efficiency than the axial version.\cite{Das2017} Finally, as suggested by de Blok \textit{et al.} for thermoacoustics,\cite{Blok2014} the possibility of scaling bidirectional turbines to the \SI{}{MW} range seems plausible. There are several examples of OWC's that have delivered more than \SI{100}{kW} of power,\cite{Das2017} with a maximum reported power output of \SI{500}{kW}.\cite{Folley2002}  

To predict the performance of the turbines, analytical calculations expressed in torque, input power, and flow coefficients can be useful for initial estimations.\cite{Setoguchi2006b,SUZUKI2008a} Furthermore, Thakker and Dnanasekaran have shown a good correspondence between a CFD model and experimental results in their study to show the large influence of the tip clearance on the turbine efficiency.\cite{Thakker2004b} However, one should be careful with using these methods blindly for thermoacoustics because they assume a quasi-steady flow. The frequency of the wave for OWC's is much smaller than the rotational frequency of the bidirectional turbine, while for thermoacoustic applications they can be of the same order. Therefore, this might mean that the calculations are sufficiently accurate for OWC's but not for thermoacoustic engines. On the other hand, in contrast with OWC's, there is a single wave frequency and amplitude in thermoacoustics. This could be used to simplify the equations for irregular bidirectional flows and one can envision turbine design optimizations that are not possible for OWC's. Therefore, the work from OWC's can be used as a starting point, but more specific research about bidirectional turbines under thermoacoustic conditions is needed to utilize its full potential.    

\section{Conclusions and recommendations\label{sec:conclusions}}
\noindent This article has given a review of the different methods to convert (thermo)acoustic power into electricity. For each technology, design aspects, operating characteristics, and methods to calculate and optimize the performance have been treated individually. This section will provide general conclusions and recommendations by the authors, which are based on a personal view and the following most important findings of the individual sections: 
\begin{itemize}
	\item Electromagnetic transducers are the most mature conversion technology in the field, and have experimentally shown the highest power output (${\sim} \SI{}{kW}$) and efficiency (${\sim} \SI{75}{\%}$). Dedicated linear alternators show the best performance and reliability, although they can be relatively expensive. Conventional loudspeakers are useful for cheap, initial prototypes, but their performance is limited and they are too fragile for long time operation.
	
	\item Piezoelectric devices provide the possibility of compact thermoacoustic engines due to their ability to operate at relatively high frequencies. The power output is generally very small (${\sim} \SI{}{mW}$), with the exception of one series of work, and the maximum efficiency of ${\sim} \SI{20}{\%}$ in thermoacoustics is not nearly as high as the ${\sim} \SI{70}{\%}$ acquired in other fields. Using dynamic magnifiers coupled with the piezoelectric material has shown to increase the output performance.      
	
	\item Magnetohydrodynamic devices have no mechanically moving parts, which makes them very reliable and suitable for inaccessible environments. The numerically predicted performance is fairly good, but there is a lack of experimental prototypes to confirm this in practice. One serious obstacle to build devices with an efficient operation is the power transduction across the gas-liquid interface.     
	
	\item Bidirectional turbines were recently introduced as potentially cheap and reliable transducers with a good performance. The efficiency in thermoacoustics can reach ${\sim} \SI{80}{\%}$ and work from the field of oscillating water columns shows that power outputs in the \SI{}{MW} range are feasible. Initial work shows an efficient performance that is nearly independent of the acoustic frequency, but more research is necessary to successfully implement bidirectional turbines in thermoacoustic engines.	
\end{itemize}

\noindent The individual conclusions show that a linear alternator is the most viable option to directly implement and efficiently convert acoustic power up to the \SI{}{kW} range. For increasing power outputs, scaling the linear alternators becomes increasingly challenging due to the large moving mass and the difficulty to maintain clearance seals over large displacement amplitudes. A promising alternative that shows no limitations in scaling the output power is the bidirectional turbine. There are good indications that its conversion efficiency can compete with the linear alternator, while the costs can be lower when relying on the mature technology of conventional rotary generators. Nevertheless, more work on bidirectional turbines needs to be done to successfully implement them and confirm that they can indeed have such a good performance. Furthermore, research should focus on optimizing the designs for the use in thermoacoustic engines, where the knowledge of a stable operating frequency and amplitude can be used to improve the turbine efficiency. As of now, piezoelectric transducers are mainly useful for designing compact thermoacoustic engines, where only a small power output is needed. One should focus on stacking multiple piezoelectric diagrams in clever designs to increase the efficiency and power output. Furthermore, to make these transducers a viable option for producing significant amounts of electricity in thermoacoustic engines, developments in the field of piezoelectric materials should lead to larger power densities and the possibility to create more bulky piezoelectric materials.

For any acoustic to electric conversion technology, it has been made clear that assuring a good power coupling between the acoustic field and the transducer is key for a good performance. An impedance mismatch will cause acoustic reflections, which results in redirected acoustic power that is not converted into electricity. Several analytical and numerical methods have been shown to be useful in ensuring a good coupling and optimizing the performance of a thermoacoustic engine, with DeltaEC being the most prominent option. The basis for the acoustic to electric transducer designs should be based on such calculations, after which prototypes can be used for further optimization. With continuing efforts on the transducers and thermoacoustic engines as a whole, the technology will be able to become a cost-effective competitor in the field of electricity production from mainly sustainable and low-grade heat sources.

\clearpage

\end{document}